\documentclass[11pt,letterpaper,notoc]{JHEP}
\usepackage{amsmath}
\usepackage{cite}
\newcommand{\SM}{ Standard Model}
\DeclareMathOperator{\tr}{tr}
\DeclareMathOperator{\diag}{diag}
\newcommand\openone{\leavevmode\hbox{\small1\normalsize\kern-.33em1}}
\providecommand{\abs}[1]{\lvert#1\rvert}

\def\CL{{\cal L}}

\def\CL{{\cal L}}

\def\ie{{\it i.e.}}

\newcommand{\gev}{ {\rm GeV} }
\newcommand{\tev}{ {\rm TeV} }

\newcommand{\lsim}{\,\raise.3ex\hbox{$<$\kern-.75em\lower1ex\hbox{$\sim$}}\,}
\newcommand{\gsim}{\,\raise.3ex\hbox{$>$\kern-.75em\lower1ex\hbox{$\sim$}}\,}

\title{The Littlest Higgs}

\author{N. Arkani-Hamed\\ Jefferson Laboratory of Physics, Harvard
  University, Cambridge, MA 02138 \\ email:
  \email{arkani@carnot.harvard.edu}}
\author{A.G.~Cohen\\ Physics Department, Boston University,
  Boston, MA 02215 \\ email:
  \email{cohen@bu.edu}}
\author{E. Katz, A.E. Nelson  \\ Department of Physics, Box 1560,
  University of Washington,  Seattle, WA 98195-1560 \\ email:
  \email{amikatz@fermi.phys.washington.edu},
  \email{anelson@phys.washington.edu}}

\preprint{UW/PT-01/07\\ HUTP-02/A017\\ BUHEP-02-23}

\abstract{We present an economical theory of natural electroweak
  symmetry breaking, generalizing an approach based on deconstruction.
  This theory is the smallest extension of the \SM\ to date
  that stabilizes the electroweak scale with a naturally light Higgs
  and weakly coupled new physics at TeV energies.  The Higgs is one of
  a set of pseudo Goldstone bosons in an $SU(5)/SO(5)$ nonlinear sigma
  model. The symmetry breaking scale $f$ is around a TeV, with the
  cutoff $\Lambda \lsim 4\pi f \sim \text{10 TeV}$.  A single
  electroweak doublet, the ``little Higgs'', is automatically much
  lighter than the other pseudo Goldstone bosons. The quartic
  self-coupling for the little Higgs is generated by the gauge and
  Yukawa interactions with a natural size ${\cal O}(g^2,\lambda_t^2)$,
  while the top Yukawa coupling generates a negative mass squared
  triggering electroweak symmetry breaking.  Beneath the TeV scale the
  effective theory is simply the minimal \SM.  The new
  particle content at TeV energies consists of one set of spin one
  bosons with the same quantum numbers as the electroweak gauge
  bosons, an electroweak singlet quark with charge 2/3, and an
  electroweak triplet scalar.  One loop quadratically divergent
  corrections to the Higgs mass are cancelled by interactions with
  these additional particles.}

\begin{document}

\section{Introduction}

The \SM\ provides an excellent effective field theory description of
almost all particle physics experiments.  But at what scale $\Lambda$
does this effective description break down? In the \SM\ the
electroweak symmetry breaking scale, of order $M_W$, is set by the
Higgs mass parameter $m_H^2$.  There is no symmetry reason why
$m^2_H/\Lambda^2$ should be small, and since this ratio receives
quantum corrections of order $\alpha_W/\pi$, a reasonable expectation
is that $\Lambda \lsim $ a few TeV \cite{'tHooft:1979bh}.
Consequently we expect new physics at the TeV scale that protects the
electroweak scale from large radiative corrections. Candidates for
this physics include technicolor, a low fundamental quantum gravity
scale and supersymmetry.  Supersymmetry is especially popular as it
provides a weakly coupled description of TeV scale physics in addition
to stabilizing a light Higgs, as seems favored by precision
electroweak data.  While these options have been vigorously explored
over the past several decades, none have yet been confirmed
experimentally.  Consequently it is important to explore qualitatively
new ideas for natural electroweak symmetry breaking, especially those
that are weakly coupled and stabilize a light Higgs.

Why should the Higgs be light?  An early
dream\cite{Georgi:1974yw,Georgi:1975tz}, dating from the seventies,
was the construction of the Higgs as a pseudo-Goldstone boson. The
first realistic attempt along these lines was made in the eighties:
the Georgi-Kaplan Composite
Higgs\cite{Kaplan:1984fs,Kaplan:1984sm,Georgi:1984af,Georgi:1984ef,Dugan:1985hq}.
In these models the Higgs is the pseudo-Goldstone boson of a
nonlinearly realized approximate global symmetry, analogous to the
pions and kaons.  Provided that $f$, the analog of the pion decay
constant, is much larger than the electroweak symmetry breaking scale,
the minimal \SM\ with a light Higgs is a good effective description of
electroweak symmetry breaking.  Unfortunately in the Georgi-Kaplan
model a hierarchy between $f$ and the electroweak scale $M_W$ is
possible only by fine tuning parameters, and so this theory does not
provide a natural stabilization of the weak scale.

In the last year, inspired by ``dimensional
deconstruction''\cite{Arkani-Hamed:2001ca,Hill:2000mu}, the Higgs as a
pseudo-Goldstone boson has been successfully
realized\cite{Arkani-Hamed:2001nc,Arkani-Hamed:2002pa}.  These models
are characterized by a ``theory space'', summarizing the gauge and
global symmetries of the theory.  The physics is described by a
nonlinear sigma model in which a special subset of the
pseudo-Goldstone bosons are naturally light, despite having gauge,
Yukawa, and self-interactions of order one\cite{Arkani-Hamed:2001nc}.

These models have a number of global symmetries, any one of which is
sufficient to ensure a massless Higgs.  These symmetries are only
approximate, explicitly broken by gauge, Yukawa and scalar couplings.
However each such coupling alone preserves enough of the global
symmetry to forbid a Higgs mass.  Quadratically divergent corrections
to the Higgs mass arise only at multi-loop order, when several such
interactions act in concert, making the small Higgs mass natural.
Such exceptionally light pseudo-Goldstone bosons were dubbed ``little
Higgses''\cite{Arkani-Hamed:2002pa}.  In a companion paper
\cite{Arkani-Hamed:2002b} we present the simplest theory space
describing a little Higgs that naturally triggers electroweak symmetry
breaking.

In this paper we show that the little Higgs phenomenon is independent
of theory space, arising in nonlinear sigma models with no obvious
nontrivial theory space description.  This allows us to realize a
little Higgs in the simplest possible way, arising from an
$SU(5)/SO(5)$ nonlinear sigma model.  The spectrum below a TeV is as
economical as possible, consisting of only the minimal \SM\
with a single light Higgs. This Higgs is a pseudo-Goldstone boson
whose decay constant $f$ is of order a TeV. At the TeV scale a small
number of additional scalars, vector bosons and quarks cancel the one
loop quadratic divergence in the Higgs mass without fine tuning or
supersymmetry.  These additional degrees of freedom represent the
smallest extension of the \SM\ to date stabilizing the weak
scale with a light Higgs and perturbative physics at TeV energies.
The cutoff of this theory can be as high as $4\pi f\sim \text{10
  TeV}$, where the nonlinear sigma model becomes strongly coupled.
The $SU(5)\to SO(5)$ symmetry breaking can easily arise from fermion
condensation through technicolor-like strong
interactions\cite{Dugan:1985hq}. However, unlike technicolor, the new
strong interactions need not appear until a scale of order 10 TeV,
with small impact on precision electroweak parameters.

\section{Requirements}

Our goal is to realize the Higgs as the pseudo-Goldstone boson of a
broken symmetry in a way which ensures that its mass is not
quadratically sensitive to the cutoff at one loop order.  This little
Higgs will then be weakly coupled up to energies one loop factor above
the weak scale, around 10 TeV.

We begin by assuming that the Higgs is part of a pseudo-Goldstone
multiplet parameterizing a coset space $G/H$, with the decay constant,
$f$, on the order of a TeV.  For phenomenological purposes, the origin
of this symmetry breaking pattern is irrelevant at energies below the
cutoff scale $\Lambda\sim 4\pi f$ where additional UV physics must
enter.  The sigma model does not Higgs the electroweak group at the
scale $\Lambda$, so the subgroup $H$ should contain $SU(2)\times
U(1)$.  As in the \SM, the electroweak gauge interactions will
na\"\i{}vely induce a one loop quadratically divergent mass for the
Higgs doublet.  To avoid this we use a mechanism familiar from models
of deconstruction: We assume that $G$ contains a weakly gauged
subgroup consisting of {\em two copies} of $SU(2)\times U(1)$: $G
\supset G_1\times G_2 = \left[SU(2)\times U(1)\right]^2$. Each of the
$G_i$ must commute with a different subgroup of $G$ that acts
non-linearly on the Higgs.  The combination of both weak gauge
interactions breaks all global symmetries which act on the Higgs, and
when both are included the Higgs ceases to be an exact Goldstone
boson.  The quadratically divergent contributions to the Higgs mass
from gauge interactions must involve both couplings, and hence first
appear at two loops.  In this case the Higgs mass squared is
radiatively stable with a cutoff of order 10 \tev.

Let us now look for a simple implementation of the above requirements.
Since $G$ contains $\left[SU(2)\times U(1)\right]^2$ it must be at
least rank 4.  Also $G$ must contain two different subgroups, each of
the form $G_i\times X_i$, $i=1,2$. Furthermore each $X_i$ must contain
an $SU(2)\times U(1)$ subgroup with some $X_i$ generators transforming
like doublets. For instance, we might (and will) take $X_i=SU(3)_i$
and $G = SU(5)$.  The doublet generators of the $X_i$ should not lie
entirely in $H$.  Assuming $G=SU(5)$, an obvious candidate for $H$ is
$SO(5)$, which contains the diagonal sum of $G_1$ and $G_2$.  With the
symmetry breaking pattern $SU(5) \to SO(5)$, the 14 Goldstone bosons
decompose under the electroweak $SU(2)\times U(1)$ as
\begin{equation}
  \mathbf{1}_0 \oplus \mathbf{3}_0 \oplus \mathbf{2}_{\pm 1/2}\oplus
  \mathbf{3}_{\pm 1}\ . 
\end{equation}
The first two sets of bosons are removed by the Higgs mechanism when
$G_1 \times G_2$ breaks to the electroweak group.  The next set are
the little Higgs and its hermitian conjugate, and the last set is an
additional complex triplet. We shall see that the triplet acquires a
TeV scale mass at one loop from gauge interactions.  The Higgs quartic
self-coupling arises from integrating out this massive triplet. Thus
the triplet coupling to the little Higgs naturally cancels the one
loop quadratic divergence in the little Higgs mass from the Higgs self
coupling.

Each of the $G_i$ gauge groups commutes with a different $SU(3)$
global symmetry subgroup of $SU(5)$.  Examining one of these $SU(3)
\times SU(2)\times U(1)$ global-local product subgroups, we see that
the first three sets of Goldstone fields above (including the little
Higgs) transform non-linearly under the $SU(3)$.  Thus neither of the
$G_i$ alone can generate a potential for the Higgs.  The two gauge
groups together however completely break all global symmetry
protecting the Higgs.  Therefore Higgs potential terms must involve
both gauge couplings and a UV sensitive Higgs mass cannot be generated
at one loop.  The triplet mass is not protected by any global symmetry
and indeed receives a quadratically cutoff sensitive mass from the
gauge interactions at one loop.  Hence below a TeV the sigma model
contains a single Higgs doublet and nothing else.  At the TeV scale
there is an additional triplet scalar, and four gauge bosons: an
electroweak triplet ${W'}^{\pm 0}$, and a neutral electroweak singlet
${B'}^0$.

\section{The Model}

Our minimal theory is based on an $SU(5)/SO(5)$ non-linear sigma
model, the same structure considered in the original Composite Higgs
models. Since this non-linear sigma model may not be as familiar as
the QCD chiral Lagrangian for pions, we describe it in some detail
here. The breaking of $SU(5) \to SO(5)$ guarantees 14 Goldstone
bosons. In order to construct the non-linear sigma model, it is
convenient to imagine for a moment that this breaking arises from a
vacuum expectation value for a $5\times 5$ symmetric matrix $\Phi$,
which transforms as $\Phi \to V \Phi V^T$ under $SU(5)$.  A vacuum
expectation value for $\Phi$ proportional to the unit matrix then
breaks $SU(5) \to SO(5)$. For later convenience, we use an equivalent
basis where the vacuum expectation value for the symmetric tensor
points in the $\Sigma_0$ direction where $\Sigma_0$ is
\begin{equation}
  \label{sigma}
  \Sigma_0=
  \begin{pmatrix} 
    \quad  & \quad & \openone \\ \quad & 1 &\quad \\ \openone
    &\quad&\quad 
  \end{pmatrix}
  \ .
\end{equation}
The unbroken $SO(5)$ generators satisfy 
\begin{align}
  T_a \Sigma_0 + \Sigma_0 T_a^T = 0
\intertext{while the broken generators obey}
  \label{brokengen}
  X_a \Sigma_0 - \Sigma_0 X_a^T = 0\ .
\end{align}
As usual, the Goldstone bosons are fluctuations about this background
in the broken directions $\Pi \equiv \pi^a X^a$, and can be
parameterized by the non-linear sigma model field
\begin{equation}
  \Sigma(x) = e^{i \Pi/f} \Sigma_0 e^{i \Pi^T/f} = 
  e^{2 i \Pi/f}  \Sigma_0,
\end{equation}
where the last step follows from \eqref{brokengen}.

We now introduce the gauge and Yukawa interactions which explicitly
break the global symmetry. As stressed in the previous section, these
are chosen to ensure an enhanced $SU(3)$ global symmetry in the limit
where any of the couplings are turned off.  We begin by gauging a $G_1
\times G_2 = [SU(2) \times U(1)]^2$ subgroup of the $SU(5)$ global
symmetry. The generators of the first $G_1 = SU(2) \times U(1)$ are
embedded into $SU(5)$ as
\begin{align}
  Q_1^a &=
  \begin{pmatrix}
    \sigma^a/2 &\quad&\quad\\ \quad&\quad&\quad
  \end{pmatrix}, &
  Y_1&= \diag (-3,-3,2,2,2)/10 \\
  \intertext{while the generators of the second $SU(2) \times U(1)$ are
    given by} 
  Q_2^a &=
  \begin{pmatrix}
    \quad&\quad&\quad\\ \quad&\quad&-{\sigma^a}^*/2
  \end{pmatrix}, &
  Y_2 &= \diag(-2,-2,-2,3,3)/10 \ .
\end{align}

In the next section, we will see that the $G_1 \times G_2$ gauge
symmetry is broken to the diagonal $SU(2) \times U(1)$ subgroup which
we identify with the electroweak gauge symmetry.  It is therefore
convenient to write the Goldstone boson matrix $\Pi$ in terms of
fields with definite electroweak quantum numbers
\begin{equation}
  \label{pgb} 
  \Pi=
  \begin{pmatrix}
    \qquad &\frac{h^\dagger}{\sqrt{2}}&\phi^\dagger\\
    \frac{h}{\sqrt{2}}&\qquad 
    &\frac{h^*}{\sqrt{2}}\\ 
    \phi&\frac{h^T}{\sqrt{2}}&\qquad
  \end{pmatrix}
\end{equation}
where $h$ is the Higgs doublet, $h=(h^+, h^0)$ and $\phi$ represents
the triplet as a symmetric two by two matrix, which transforms as a
$\mathbf{3}_1$ under the electroweak group.  We have ignored the
Goldstone bosons that are eaten in the Higgsing of $[SU(2)\times
U(1)]^2 \to SU(2)\times U(1)$.

These gauge interactions satisfy the requirements of the previous
section. When the couplings of $G_1$ are turned off, there is an
enhanced $SU(3)_1$ global symmetry living in the upper $3 \times 3$
block of $SU(5)$; when the gauge interactions of $G_2$ are turned off
there is an $SU(3)_2$ global symmetry living in the lower $3 \times 3$
block.

The tree-level Lagrangian for the model is given by 
\begin{equation}
  \label{eq:lag}
  {\cal L} = {\cal L}_K + {\cal L}_t + {\cal L}_\psi
\end{equation}
Here ${\cal L}_K$ contains the kinetic terms for all the fields;
${\cal L}_t$ generates the top Yukawa coupling; and ${\cal L}_\psi$
generates the remaining small Yukawa couplings. In detail, ${\cal
  L}_K$ includes the conventional kinetic terms for the gauge fields
and Fermions, as well as the leading two-derivative term for the
non-linear sigma model
\begin{equation}
  \frac{f^2}{4} \tr\abs{D_\mu\Sigma}^2 
\end{equation}
where the covariant derivative of $\Sigma$ is given by 
\begin{equation}
  \label{eq:cd}
  D\Sigma = \partial \Sigma - \sum_j \left\{
   i g_{j} W_j^a (Q_j^a \Sigma + \Sigma Q_j^{aT}) 
  + i g'_{j} B_j (Y_j \Sigma + \Sigma Y_j^T)\right\}.
\end{equation}
The $g_i, g^\prime_i$ are the couplings of the $\left[SU(2)\times
  U(1)\right]_i$ groups. In order to introduce a large top Yukawa
coupling while avoiding the associated large quadratic divergence in
the Higgs mass, we add a pair of colored Weyl Fermions
$\tilde{t},\tilde{t}^c$ in addition to the usual third-family weak
doublet $q_3 = (t_3,b_3)$ and weak singlet ${u^\prime_3}^c$. It is
convenient to group the doublet together with
$\tilde{t}$ into a row vector $\chi = (b_3 \, t_3 \, \tilde{t})$. ${\cal
  L}_t$ is given by
\begin{equation}
  \label{topmass1}
  \CL_{t}=\lambda_1 f  
  \epsilon_{ijk} \epsilon_{xy}
  \chi_i \Sigma_{jx}\Sigma_{ky} {u^\prime_3}^c+
  \lambda_2 f \tilde{t} \tilde{t}^c + \text{h.c.}\ ,
\end{equation}
where the indices $i,j,k$ are summed over $1,2,3$ and $x,y$ are summed
over $4,5$. This interaction fulfills our requirements: the
$\lambda_1$ interaction preserves the $SU(3)_1$ and breaks $SU(3)_2$,
while $\lambda_2$ does the converse. To see that ${\cal L}_t$
generates a top Yukawa coupling we expand ${\cal L}_t$ to first order
in the Higgs $h$:
\begin{equation}
  {\cal L}_t = \lambda_1  (q_3 h + f \tilde{t}) {u^\prime_3}^c  +
  \lambda_2 f \tilde{t} \tilde{t}^c + \cdots\ .
\end{equation}
Clearly $\tilde{t}$ marries one linear combination of ${u^\prime_3}^c$
and $\tilde{t}^c$ to become massive. Integrating out this heavy quark,
the remaining combination $u_3^c$ has the desired Yukawa coupling to
$q_3$ 

\begin{equation}
  \label{topyuk}
  \lambda_t \, q_3 h \, u_3^c, \qquad \text{where} \qquad
  \lambda_t = \frac{\lambda_1 
    \lambda_2}{\sqrt{\lambda_1^2 + \lambda_2^2}}\ .
\end{equation}
 
The mixing of the top quark with vector-like Fermions at the
TeV scale is similar to Frogatt-Nielsen models of flavor
\cite{Froggatt:1979nt} and the top see-saw \cite{Dobrescu:1998nm,
  Chivukula:1998wd}.
Finally, the interactions in ${\cal L}_\psi$ encode the remaining
Yukawa couplings of the \SM. These couplings are small so that the
1-loop quadratically divergent contributions to the Higgs mass they
induce are negligible with a cutoff $\Lambda \sim \text{10 TeV}$.  For
the up sector we can take ${\cal L}_\psi$ to have exactly the same
form as ${\cal L}_t$, except that the $\tilde{t},\tilde{t}^c$ fields
are unnecessary. For the down and charged lepton sector we use the
same Lagrangian with $\Sigma$ replaced by $\Sigma^*$.

The $U(1)$ charges of the Fermions are chosen to ensure gauge
invariance. As we will see in the next section, the $G_1 \times G_2$
symmetry is Higgsed to the diagonal \SM\ $SU(2) \times U(1)$ gauge
group, so the $U(1)$ charges must be chosen to yield the correct \SM\ 
quantum numbers.  We do not concern ourselves with cancellation of the
$G_1\times G_2$ anomalies in this low energy effective theory, since
there may be additional Fermions at the cutoff which cancel the
anomalies involving the broken subgroup. We insist only that \SM\ 
anomalies cancel, since fermions in a chiral representation of the
\SM\ can have only weak scale masses. Notice also that our
two $U(1)$ generators $Y_1,Y_2$ are not orthogonal, but we have not
included any kinetic mixing term between them. Any mixing between the
$U(1)$'s in our effective theory arises at loop level and is
sufficiently small. In this model our choice for the $U(1)$'s is
dictated by the requirement that in the limit where any one of the
$U(1)$'s is turned off there is an $SU(3)$ global symmetry. If we
demand that the two $U(1)$'s {\em both} fit inside the $SU(5)$ global
symmetry, the $Y_1,Y_2$ are fixed. It is easy to add new $U(1)$
factors that commute with the $SU(5)$ global symmetry which allow
$Y_1,Y_2$ to be chosen orthogonal. Alternatively, enlarging to an
$SU(N)/SO(N)$ non-linear sigma model allows sufficient room to embed
two orthogonal $U(1)$'s in the $SU(N)$ global symmetry while still
satisfying our requirements.

\section{Radiative corrections and electroweak symmetry breaking}

Since the gauge and Yukawa interactions explicitly break the global
$SU(5)$ symmetry, these interactions will select a preferred alignment
for $\Sigma$.  To find the vacuum, we compute the 1-loop radiative
potential for $\Sigma$ from the gauge, scalar and Yukawa sectors.

First consider the gauge sector. The largest corrections come
from 1-loop gauge quadratic divergences, which are easily extracted
from the quadratically divergent part of the Coleman-Weinberg
potential\cite{Coleman:1973jx}:
\begin{equation}
  \frac{\Lambda^2}{16 \pi^2} \tr M_V^2(\Sigma)
\end{equation}
where $M^2(\Sigma)$ is the gauge boson mass matrix in a background
$\Sigma$. $M_V^2(\Sigma)$ can be easily read off from the covariant
derivative for $\Sigma$ \eqref{eq:cd}, giving a potential
\begin{equation}
  \label{eq:cw}
  c g^2_{j} f^4 \sum_a \tr \left[(Q_j^a \Sigma)(Q_j^a \Sigma)^*\right]
  + c {g'_j}^2 f^4 \tr \left[(Y_j \Sigma)(Y_j \Sigma)^*\right]
\end{equation}
Here we have used $\Lambda \sim 4 \pi f$, and $c$ is an ${\cal O}(1)$
constant whose precise value is sensitive to the UV physics at the
scale $\Lambda$.  The presence of this quadratic divergence implies
that the Lagrangian \eqref{eq:lag} is incomplete: we must include all
operators consistent with the symmetries of the theory with natural
sizes determined by na\"\i{}ve dimensional
analysis\cite{Manohar:1984md,Cohen:1997rt,Luty:1998fk}.  At second
order in the gauge couplings and momenta \eqref{eq:cw} is the unique
gauge invariant term transforming properly under the global $SU(5)$
symmetry.  This potential is analogous to that generated by
electromagnetic interactions in the pion chiral Lagrangian, which
shift the masses of $\pi^\pm$ from that of the $\pi^0$
\cite{Das:1967it}. For the pion masses, na\"\i{}ve dimensional
analysis works beautifully.  Moreover, in analogy to the chiral
Lagrangian, we assume that $c$ is positive. This implies that the
vacuum is correctly aligned, the electroweak group remains unbroken by
the sigma model, and the triplet has a positive TeV sized mass $m_\phi
\sim g f$.

To find the Higgs potential we expand the Lagrangian \eqref{eq:cw} in
the pseudo-Goldstone fields.  The form of the potential is determined
by the global symmetry transformation properties of the Higgs and
triplet fields.  The $G_1$ gauge interactions leave the $SU(3)_1$
symmetry invariant, part of which acts on the Higgs and triplet fields
as
\begin{align}
  h_i &\rightarrow  h_i + \epsilon_i + \cdots \\
  \phi_{ij} &\rightarrow \phi_{ij} - i(\epsilon_i h_j + \epsilon_j
  h_i) + \cdots  
\intertext{while $G_2$ leaves $SU(3)_2$ symmetry invariant, and acts as}
  h_i & \rightarrow h_i + \eta_i + \cdots \\
  \phi_{ij} &\rightarrow \phi_{ij} + i(\eta_i h_j + \eta_j
  h_i) + \cdots\ .
\end{align}
Hence to quadratic order in $\phi$ and quartic order in $h$ the
potential is
\begin{equation}
  \label{eq:hpot}
  c (g_1^2+{g_1'}^2) f^2 \abs{\phi_{ij} + \frac{i}{2f} (h_i
    h_j  + h_j h_i)}^2
  + c ( g_2^2+ {g_2'}^2) f^2 \abs{\phi_{ij} - \frac{i}{2f} (h_i h_j +
    h_j h_i)}^2
\end{equation}
As previously claimed the gauge interactions induce a mass for the
triplet of order $g f$, while the little Higgs remains massless.

There is also a quadratically divergent Coleman-Weinberg potential
generated by the Fermion loop, which requires the inclusion of the
operator
\begin{equation}
  \label{topct}
  - c'\lambda_1^2 \epsilon^{wx}\epsilon_{yz} \epsilon^{ijk}\epsilon_{kmn}
  \Sigma_{iw}\Sigma_{jx}
  \Sigma^{*my}\Sigma^{*nz}+ \text{h.c.}
\end{equation}
This operator is $SU(3)_1$ symmetric, and therefore generates a
potential of the same form as the first term in \eqref{eq:hpot}, with
coefficient $-c' \lambda_1^2$. As long as $c(g_1^2 + {g_1^\prime}^2 +
g_2^2 + {g_2^\prime}^2) - c^\prime \lambda_1^2 > 0$, the triplet mass
squared remains positive. At energies beneath the triplet mass, we can
integrate this particle  out and get a quartic potential for the Higgs
\begin{equation}
  \label{eq:1}
  \lambda (h^\dagger h)^2, \quad \text{where} \quad \lambda = c
  \frac{(g_1^2 +{g_1^\prime}^2 - c^\prime/c \lambda_1^2)(g_2^2
    +{g_2^\prime}^2)} 
  {g_1^2 +{g_1^\prime}^2 - c^\prime/c \lambda_1^2 + g_2^2
    +{g_2^\prime}^2} 
\end{equation}
As advertised the interactions combine to give the Higgs a quartic
potential determined by gauge and Yukawa couplings, and no mass term.

The remaining part of the vector boson contribution to the
Coleman-Weinberg potential is
\begin{equation}
  \frac{3}{64\pi^2}\tr
    M_V^4(\Sigma)\log
      \frac{M_V^2(\Sigma)}{\Lambda^2}\ .
\end{equation} 
This gives a logarithmically enhanced positive Higgs mass squared
\begin{equation} 
\label{eq:vector}
  \frac{3}{64\pi^2} \left\{3
    g^2 {M'_W}^2\log\frac{\Lambda^2}{{M'_W}^2}+{g'}^2
    {M'_B}^2\log
      \frac{\Lambda^2}{{M'_B}^2}\right\}
\end{equation} 
where $M'_W$ is the mass of the heavy $SU(2)$ triplet of gauge bosons
and $M'_B$ is the mass of the heavy $U(1)$ gauge boson.

There is a similar Coleman-Weinberg potential from the scalar
self-interactions in \eqref{eq:cw} and in \eqref{topct} which also
give logarithmically enhanced positive contributions to the Higgs mass
squared:
\begin{equation}
  \label{eq:2}
  \frac{\lambda}{16\pi^2}M_\phi^2
  \log\frac{\Lambda^2}{M_\phi^2}
\end{equation}
where $M_\phi$ is the triplet scalar mass.

The remaining part of the Fermion loop contribution to the
Coleman-Weinberg potential is
\begin{equation}
     - \frac{3}{16\pi^2} \tr \left(M_f(\Sigma)
      M_f^\dagger(\Sigma)\right)^2\log\frac{M_f(\Sigma)
          M_f^\dagger(\Sigma)}{\Lambda^2}
\end{equation}
where $M_f(\Sigma)$ is the fermion mass matrix in a background
$\Sigma$.  This potential gives a logarithmically enhanced, {\em
  negative} contribution to the Higgs mass squared
\begin{equation}
  \label{eq:3}
     -\frac{3\lambda_t^2}{8 \pi^2}
     {m'}^2\log\frac{\Lambda^2}{{m'}^2}
\end{equation}
where $m'$ is the mass of the heavy fermion. This can dominate over
the positive gauge and scalar contributions, triggering electroweak
symmetry breaking.

What is the mass of the physical Higgs in this model? The Higgs mass
is determined by the Higgs quartic coupling $\lambda$, which
receives significant contributions from the gauge interactions
\eqref{eq:1} and from the operator \eqref{topct}. Both of these
contributions are proportional to unknown coefficients $c, c'$ of
order one, encoding information about the UV physics. We can obtain a
more predictive theory for the Higgs mass through an alternative model
for the top Yukawa coupling.  We introduce fermions in complete
$SU(5)$ multiplets, transforming as $(5,3)$ and $(5, \bar 3)$ under
$SU(5)\times SU(3)_{\text{color}}$ and coupling to the $\Sigma$ field
in an $SU(5)$ symmetric fashion.  Such multiplets might be expected in
strongly coupled theories. The left handed top and bottom are a
mixture of a component of a $(5,3)$ multiplet and an additional quark
doublet field $q\sim(t,b)$ and the anti-top is a similar mixture of a
component of the $(5,\bar 3)$ field and an $SU(5)$ singlet field
$t^c$. We break the $SU(5)$ symmetry only through explicit fermion
mass terms connecting the $q$ and $t^c$ to the $SU(5)$ multiplet
fermions with the appropriate quantum numbers.  This form of symmetry
breaking is soft enough to not induce quadratic divergences at one
loop, and so the gauge contribution dominates the Higgs quartic
potential. In this case the Higgs mass is parametrically of order the
$Z$ mass, $m_H \sim M_Z$!

\section{Precision Electroweak tests}
\label{precision}

New physics which couples to the Higgs and gauge bosons is constrained
by precision electroweak measurements, which agree well with the
predictions of the minimal \SM.  The contributions of new, weakly
coupled particles are generally suppressed by factors of
$M^2_W/M^2_{\text{new}}$, unless the mass $M_{\text{new}}$ of the new
particles is due to electroweak symmetry breaking, in which case the
contributions of heavy particles to low energy physics do not decouple
and can be large, as occurs in technicolor
theories\cite{Holdom:1990tc,Peskin:1990zt,Golden:1991ig}. In our case
all the new couplings are weak and all the new particle masses are
symmetric under the electroweak interactions, hence their loop
contributions are of order $(g^2/(16 \pi^2)) M^2_W/(g f)^2$, \ie \ of
similar size to 2-loop \SM\ corrections. Corrections of this size are
typically smaller than the current experimental uncertainties.  The
main effects on low energy observables will arise from tree-level
exchange of heavy particles, which give effects 
comparable to \SM\ loops. For instance the $\phi$ scalar
will have a custodial $SU(2)$ breaking trilinear coupling with the
Higgs of the form $ h\phi h$. Integrating out the $\phi$ at tree level
will induce a custodial $SU(2)$ violating dimension six operator,
which contributes to the $T$ (or $\rho$) parameter.  Fortunately, the
coefficient of the trilinear coupling is not expected to be large. The
gauge contribution is proportional to $g_1^2-g_2^2+{g'_1}^2-{g'_2}^2$
while the Fermion contribution is proportional to $-c^\prime
\lambda_1^2$. This would be absent in the complete $SU(5)$ multiplet
model of the top mass.  Another source of tree level corrections is
due to the exchange of the heavy gauge bosons, which have tree-level
couplings to Fermions. For instance, $W'$ exchange can contribute to
muon decay and affect the muon lifetime.  Similar studies of the
contributions of heavy gauge bosons to precision electroweak
corrections give lower bounds on the masses as tight as 3 TeV
\cite{Marciano:1999ih}.  The bounds would be weaker in the present
case if, for instance, the two $SU(2)$'s have unequal couplings and
the light fermions only transform under the more weakly coupled of the
two $SU(2)$ gauge groups.  We leave a study of the effects on
precision electroweak observables for future work.

\section{UV Completions and the Resurrection of Strong Dynamics for EWSB}

So far, we have contented ourselves with an effective theory
description of our non-linear Sigma model beneath the scale $\Lambda
\sim 10$ TeV, but it is interesting to contemplate a UV origin for our
physics. There is a straightforward UV completion into a linear sigma
model, suitably supersymmetrized to avoid quadratic divergences coming
from above the scale $\Lambda$, with a supersymmetry breaking scale
$\sim 100$ TeV. But it is fascinating that the essential features of
our model can arise from strong gauge dynamics at the scale $\Lambda$.
Consider, an $SO(N)$ gauge theory with $5$ Weyl Fermions $\Psi_i$ in
the fundamental representation of $SO(N)$. When the $SO(N)$ coupling
becomes strong, the Fermions condense as
\begin{equation}
  \langle \Psi_i \Psi_j \rangle \sim \Lambda^3 \Sigma_{ij}
\end{equation}
breaking $SU(5) \to SO(5)$, where $\Lambda$ is the $SO(N)$ scale and
the $\Sigma_{ij}$ parameterize the Goldstone Bosons in an
$SU(5)/SO(5)$ non-linear Sigma model field. Just as in our effective
Lagrangian description, we can now weakly gauge a subgroup of the
$SU(5)$ global symmetry, and at low energies, we have a little Higgs
with an an exceptionally light mass $m_H \sim (g^2 \Lambda/16 \pi^2)$,
even though it has large gauge coupling and a large quartic
self-coupling of order $g^2$.  Unfortunately, because $Y_1,Y_2$ are
not orthogonal, there will be logarithmically enhanced kinetic mixing
between them from a $\Psi$ loop, which induces 1-loop quadratic
divergences for the little Higgs mass from the $U(1)$ interactions.
This minor annoyance can be avoided by slightly enlarging the model.
For example, $7$ rather than $5$ Weyl Fermions leads to an
$SU(7)/SO(7)$ model. In this case we can weakly gauge a $G_1 \times
G_2 = [SU(2) \times U(1)] \times SU(3)$ subgroup of the $SU(7)$ global
symmetry, with the $U(1)$ embedded as $Y =\diag(1,1,-2,0,0,0,0)$.
These charges satisfy our requirements: the $G_1$ couplings preserve
an $SU(4)$ global symmetry while the $G_2$ coupling preserves a
different $SU(4)$ global symmetry. There is a triplet of little
Higgses in this model, consisting of a conventional \SM\ Higgs and
charged singlet.

For a realistic model we still need to incorporate Fermion masses,
most importantly the large top Yukawa coupling to drive electroweak
symmetry breaking. Therefore an ETC-like mechanism is still likely
needed to generate operators of the form of equation \eqref{topmass1}.

The construction of a little Higgs from strong dynamics resurrects the
possibility that strong dynamics play a role in EWSB while avoiding
the traditional pitfalls of technicolor, and greatly ameliorating the
difficulties of extended technicolor. Technicolor and ETC are beset by
three major woes: excessively large correction to precision
electroweak observables, such as the $S$-parameter; excessively light
pseudo-Goldstone bosons with electroweak quantum numbers; and large
flavor-changing neutral currents.  Little Higgses eliminate the first
two problems: electroweak symmetry breaking is accomplished with a
light weakly coupled Higgs, and there are no large corrections to the
$S$-parameter. The pseudo-Goldstone boson issue is used to advantage:
our little Higgs {\em is} the only excessively light pseudo-Goldstone
boson with electroweak quantum numbers! As for FCNCs, the higher
scale for the strong dynamics significantly relaxes the constraints,
as we discuss in the next section.

\section{Flavor Changing Neutral Currents}

Strictly within our low-energy effective field theory, we do not
have to worry about FCNC. The only flavor-violating interactions
beyond those in the SM Yukawa couplings involve the top sector, and
therefore there are no large FCNCs involving the light fermions.
However, in some UV completions, other, less benign spurions might be
present. In order to address FCNC issues we need to speculate about
the UV origin of the spurions which give the quark and lepton
couplings to the $\Sigma$ field.

As stressed in \cite{Chivukula:2002ww}, in those UV completions of
the deconstruction models in which the light quark
couplings to the sigma field arise from four Fermi interactions in the
UV theory, there may also be four Fermi interactions among light
quarks, with coefficients proportional to their Yukawa couplings.  If
the light quark four Fermi couplings are about the same size as the
couplings of the light quarks to the new fermions, then for
$\Lambda\lsim 75$ \tev, some of the resulting FCNCs
could exceed the experimental bounds.  Also
discussed in \cite{Chivukula:2002ww} were ways to
ameliorate these effects, such as anomalous scaling.

Here we briefly mention some other scenarios in which one could
explain the observed suppression of FCNCs.  If the sigma field is a
condensate of strongly interacting fermions, and so our little Higgs
is composite, one might imagine that the fermions are also composite,
with masses protected by approximate global symmetries
\cite{'tHooft:1979bh}.  It is then a simple matter to postulate global
symmetries and spurions which allow the necessary Yukawa couplings
while preventing excessive FCNC \cite{Chivukula:1987py}.  Another
possibility is that the world is supersymmetric, with supersymmetry
broken at 100 TeV or so. Then scalars with masses of 100 TeV are
natural.  If strongly coupled to the constituents giving rise to the
nonlinear sigma fields and weakly coupled to light quarks and leptons,
100 TeV scalars could mediate interactions leading to acceptable
fermion masses without excessive FCNCs
\cite{Simmons:1989fu,Samuel:1990dq,Dine:1990jd}.

\section{Smoking Guns}

We will not attempt an analysis of the experimental consequences of
this model here, but simply mention qualitative features of the most
distinctive signals. Unlike in solutions to the hierarchy problem
involving extra dimensions, deconstructed extra dimensions, or
supersymmetry, the natural expectation for the mass of all new
particles beyond the \SM\ is of order a TeV.  Unlike in
technicolor, there is a weakly coupled light Higgs and no new strong
interactions below the cutoff. There are, however, several unique
signatures for the discovery of new particles at the LHC. The most
unique feature of the $SU(5)/SO(5)$ nonlinear sigma model is the
$\phi$ electroweak triplet scalar, which will appear as three nearly
degenerate scalars with charges $2,1$ and $0$.  For aspects of the
phenomenology see \cite{Georgi:1985nv,Chivukula:1986sp,Gunion:1996pq}.
The other distinctive features of this model, shared with
deconstruction models of electroweak symmetry breaking, are the
additional $SU(2)\times U(1)$ gauge bosons and charge 2/3 quarks.

Without specification of the UV physics above the cutoff, the Higgs
potential and couplings are determined at one loop in terms of 9
parameters of the effective theory below the cutoff---these are
$g_{1,2},g'_{1,2},\lambda_{1,2},f,c$ and $c'$. Four combinations of
these are determined from $\alpha$, $\sin^2\theta_W$, the top mass,
and the Higgs vev, while the other five could be determined from the
Higgs mass, $m'$, $M_W'$, $M_B'$, and $M_\phi$, or in principle, given
sufficient precise data, from a fit to electroweak observables.  It is
possible that the latter could give a prediction for the masses of the
new particles.

We have chosen to eliminate all 1-loop quadratic divergences in the
Higgs mass squared parameter. The apparent divergences in the low
energy effective theory from the \SM\ gauge bosons, top quark and Higgs
scalar are cancelled by new particles of similar statistics.  As the
masses of these new particles increase, the difference between the \SM\
1-loop contributions and that of these new particles grows, requiring
fine tuning of the Higgs mass squared parameter.  Hence the masses of
these new particles may be bounded by requiring that this cancellation
is not finely tuned. The cutoff dependence may be absorbed into the
counterterms $c$ and $c'$.  The logarithm might be as large as
$\log(16\pi^2)\sim 5$, however to be conservative we will take it to
be 1.
The reasonable naturalness constraint that none of the independent
contributions in \eqref{eq:vector},\eqref{eq:2},\eqref{eq:3} exceed
the absolute value of the Higgs mass squared parameter by more than a
factor of 10 ($\sim10$\% fine-tuning) gives upper bounds on the new
particle masses as a function of the physical Higgs mass. The new charge 2/3
quark is the most constrained:
\begin{equation} 
  m'\lsim 2 \ \tev\ \left(\frac{m_H}{200\ \gev}\right)^2.
\end{equation}
Two dominant decay modes are the flavor changing neutral current
$T'\rightarrow Z t$, due to the mass mixing of charge 2/3 quarks with
different weak charges, and $T' \rightarrow h t$.  Using the
expression \eqref{topyuk} for the top Yukawa coupling, we conclude
that $\sqrt{\lambda_1^2 + \lambda_2^2} > 2 \lambda_t$. Since the mass
of the heavy quark is $m' = \sqrt{\lambda_1^2+\lambda_2^2}\ f$, the
naturalness bound on $m'$ in turn implies
\begin{equation}
  \label{fbound}
  f\lsim 1\ \tev\ \left(\frac{m_H}{200\ \gev}\right)^2.
\end{equation}
Given the expectation that the couplings are all weak, \eqref{fbound}
suggests that all the new particles should have masses around a few
TeV and are available for an LHC discovery. However the 1-loop
naturalness bounds on the new bosons are not stringent enough to be
interesting.  These are:
\begin{equation} 
  M_W' \lsim   6\ \tev \left(\frac{m_H}{200\ \gev}\right)^2 ,\quad
  M_\phi \lsim 10\ \tev
\end{equation}
One might try to obtain tighter bounds from estimating 2-loop
contributions, but these remain quadratically sensitive to the cutoff and
thus constrain the cutoff physics rather than the parameters of the
effective theory.

Although not guaranteed by naturalness, discovery of the new particles
at the Tevatron run II is not out of the question.

\section{Conclusions}

Theories with a little Higgs---where the lightness of the Higgs is
understood because it is a pseudo-Goldstone Boson---provide a
qualitatively new framework for physics beyond the \SM.  While the
first examples of such models were inspired by deconstruction and
theory space, in this paper we have seen how these ideas can be
generalized to yield very economical models.  The essential
requirement is that the Higgs should transform nonlinearly under a
collection of symmetries, which are completely broken by a collection
of spurions, but no single spurion should break all the symmetries.
We exploited this insight to present what we believe is the minimal
possible set of new symmetries and particles needed to stabilize the
weak scale against a cutoff of order $\Lambda \sim 10 \text{TeV}$,
without fine tuning. We have logarithmic sensitivity to the cutoff at
one loop, and quadratic sensitivity at 2-loops, which is sufficient to
make the electroweak symmetry breaking scale of 250 GeV natural. Our
philosophy here is rather similar to that of Effective Supersymmetry
\cite{Dimopoulos:1995mi,Cohen:1996vb} in which only the minimal set of
superpartners required for naturalness is kept lighter than the TeV
scale, with all others pushed up to 10 TeV, but our particle content
at the TeV scale is much more economical.

We could eliminate all UV sensitivity to some specified number of
loops, and thereby obtain more predictivity, at the price of being 
less minimal---introducing larger
symmetry groups and more particles---as was done in
\cite{Arkani-Hamed:2001nc,Arkani-Hamed:2002pa}.  For now, however,
there is no experimental motivation to do so, as a 10 TeV cutoff is
sufficient to account for the agreement of the \SM\ with
precision electroweak data.

This model is the simplest example in a new class of theories of
natural electroweak symmetry breaking, and clearly its phenomenology
deserves further study. There are many avenues for further
exploration, including generalizations beyond our minimal model,
calculations of precision electroweak observables, as well as possible
UV completions.

Our minimal model is remarkable in providing the first example of a
theory of natural electroweak symmetry breaking with {\em no} new
degrees of freedom beyond the \SM\ beneath a TeV. This is in sharp
contrast to the MSSM, where there is no reason for the Higgs to be
lighter than the superpartners. Even above a TeV, our model introduces
only a very small number of new degrees of freedom that stabilize the
Higgs mass. Counting all helicity states, the triplet scalar, massive
gauge bosons and heavy fermion add a total of 30 new real degrees of
freedom beyond the \SM. This is smaller than the 56 new
degrees of freedom introduced in the minimal moose of our companion
paper \cite{Arkani-Hamed:2002b}, and far smaller than the 126 new
degrees of freedom introduced in the MSSM, not to mention the $\sim
1000$ new degrees of freedom introduced in theories with extra
dimensions at the TeV scale. Of course, mindless minimalism is not a
measure by which to judge a physical theory, but the economy of our
model does illustrate the simplicity of the underlying mechanism.

Summarizing, the broad consequences of our model are threefold:
\begin{enumerate}
\item Electroweak symmetry breaking without fine tuning can be
  realized with the particle content of the minimal \SM\
  below a TeV.
\item A small number of new, weakly coupled particles are required at
  a few TeV, including at least one heavy copy of the electroweak
  gauge bosons and top quark, and a scalar coupled to the Higgs.
\item The old idea of dynamical electroweak symmetry breaking can be
  resurrected, naturally manifesting itself at low energies as the
  \SM\ with a weakly coupled light Higgs.

\end{enumerate}

\acknowledgments 
We would like to thank Sekhar Chivukula, Ken Lane and Martin Schmaltz for 
stimulating conversations. The work of A. Nelson and E. Katz was partially
supported by the DOE under contract DE-FGO3-96-ER40956.  A.G.C. is
supported in part by the Department of Energy under grant number
DE-FG02-91ER-40676.  N.A-H. is supported in part by the Department of
Energy under Contracts DE-AC03-76SF00098 and the National Science
Foundation under grant PHY-95-14797, the Alfred P. Sloan foundation,
and the David and Lucille Packard Foundation.



\providecommand{\href}[2]{#2}\begingroup\raggedright\endgroup

\end{document}